\documentclass[prb,superscriptaddress,floatfix,letterpaper,twocolumn,aps,showpacs]{revtex4-1}
\usepackage[utf8]{inputenc}
\usepackage{natbib}
\usepackage{amsmath}
\usepackage{amsbsy}
\usepackage{amssymb}
\usepackage{scrextend}
\usepackage{graphicx}
\usepackage{color}
\usepackage{xcolor}

\newcommand{\eq}{Eq.\ }

\newcommand{\sbonlinecite}[1]{[\onlinecite{#1}]}

\renewcommand{\vec}[1]{\boldsymbol{#1}}
\newcommand{\ooff}{$\mathrm{Rb}\mathrm{Eu}\mathrm{Fe}_{4}\mathrm{As}_{4}$}
\newcommand{\BKT}{\mathrm{BKT}}

\newcommand{\sss}{\scriptscriptstyle}
\newcommand{\kB}{k_{\sss B}}
\newcommand{\muB}{\mu_{\sss B}}

\newcommand{\cpi}{\cos\phi_{i}}
\newcommand{\spi}{\sin\phi_{i}}
\newcommand{\cti}{\cos\theta_{i}}
\newcommand{\sti}{\sin\theta_{i}}

\newcommand{\ctj}{\cos\theta_{j}}
\newcommand{\stj}{\sin\theta_{j}}

\newcommand{\llangle}{\langle\!\langle}
\newcommand{\rrangle}{\rangle\!\rangle}

\begin{document}
\title{Strongly-fluctuating moments in the high-temperature magnetic superconductor $\mathrm{Rb}\mathrm{Eu}\mathrm{Fe}_{4}\mathrm{As}_{4}$}
\author{K.\ Willa}
\affiliation{Materials Science Division, Argonne National Laboratory, 9700 South Cass Avenue, Lemont, IL 60439, USA}
\author{R.\ Willa}
\affiliation{Materials Science Division, Argonne National Laboratory, 9700 South Cass Avenue, Lemont, IL 60439, USA}
\author{J.-K.\ Bao}
\affiliation{Materials Science Division, Argonne National Laboratory, 9700 South Cass Avenue, Lemont, IL 60439, USA}
\author{A. E.\ Koshelev}
\affiliation{Materials Science Division, Argonne National Laboratory, 9700 South Cass Avenue, Lemont, IL 60439, USA}
\author{D. Y.\ Chung}
\affiliation{Materials Science Division, Argonne National Laboratory, 9700 South Cass Avenue, Lemont, IL 60439, USA}
\author{M. G.\ Kanatzidis}
\affiliation{Materials Science Division, Argonne National Laboratory, 9700 South Cass Avenue, Lemont, IL 60439, USA}
\affiliation{Department of Chemistry, Northwestern University, Evanston, Illinois, 60208, USA}
\author{W.-K.\ Kwok}
\affiliation{Materials Science Division, Argonne National Laboratory, 9700 South Cass Avenue, Lemont, IL 60439, USA}
\author{U.\ Welp}
\affiliation{Materials Science Division, Argonne National Laboratory, 9700 South Cass Avenue, Lemont, IL 60439, USA}

\date{\today}

\begin{abstract}
The iron-based superconductor $\mathrm{Rb}\mathrm{Eu}\mathrm{Fe}_{4}\mathrm{As}_{4}$ undergoes a magnetic phase transition deep in the superconducting state. We investigate the calorimetric response of $\mathrm{Rb}\mathrm{Eu}\mathrm{Fe}_{4}\mathrm{As}_{4}$ single crystals of the magnetic and the superconducting phase and its anisotropy to in-plane and out-of-plane magnetic fields. Whereas the unusual cusp-like anomaly associated with the magnetic transition is suppressed to lower temperatures for fields along the crystallographic $c$ axis, it rapidly transforms to a broad shoulder shifting to higher temperatures for in-plane fields. We identify the cusp in the specific heat data as a Berezinskii-Kosterlitz-Thouless (BKT) transition with fine features caused by the three-dimensional effects. The high-temperature shoulder in high magnetic fields marks a crossover from a paramagnetically disordered to an ordered state. This observation is further supported by Monte-Carlo simulations of an easy-plane 2D Heisenberg model and a fourth-order high-temperature expansion; both of which agree qualitatively and quantitatively with the experimental findings.
\end{abstract}

\maketitle
While superconductivity and magnetic order usually are mutually exclusive due to their competitive nature, a series of novel materials that feature the coexistence of both phases has recently emerged \cite{Muller2001, Kulic2008, Zapf2017}. In order to address open questions on the coexistence/interplay/competition between these two phases of matter, it is crucial to study model systems, where both phenomena can be tuned independently from each other. The $\mathrm{Eu}$-based pnictide superconductors---where superconductivity occurs within the $\mathrm{Fe}_{2}\mathrm{As}_{2}$ layers, while the magnetism is hosted by the $\mathrm{Eu}^{2+}$ ions---provides such a model system \cite{Zapf2017}. Furthermore, each phenomenon appears to be relatively robust against perturbing the other one. In fact, chemical substitution of the parent non-superconducting compound $\mathrm{Eu}\mathrm{Fe}_{2}\mathrm{As}_{2}$, e.g. with $\mathrm{P}$ (on the $\mathrm{As}$ site), $\mathrm{K}$ or $\mathrm{Na}$ (on the $\mathrm{Eu}$ site) induces superconductivity \cite{Cao2011, *Jeevan2011, Jeevan2008, Qi2008} [with maximum $T_{c}$ of $23\mathrm{K}$, $30\mathrm{K}$, and $35\mathrm{K}$ respectively], while only smoothly suppressing the magnetic order temperature $T_{m} \!\sim\! 19\mathrm{K}$. Recent syntheses \cite{Liu2016a, Liu2016b} of members of the 1144 family ($\mathrm{Cs}\mathrm{Eu}\mathrm{Fe}_{4}\mathrm{As}_{4}$ and {\ooff} with $T_{c}$ in the mid-$30\mathrm{K}$ range) have opened new possibilities to tune the separation, and hence the interaction between neighboring $\mathrm{Eu}$ layers.

In this paper we report a detailed calorimetric characterization of single crystal \ooff: in particular, we investigate the anisotropic response near the magnetic phase transition at $T_{m} = 14.9\mathrm{K}$ (well within the superconducting state, $T_{c} = 37\mathrm{K}$) to external fields. Whereas earlier studies on polycrystalline samples \cite{Liu2016b} have 
suggested that the magnetic transition might be of third (higher-than-second) order, we demonstrate that the behavior of the specific heat is broadly consistent with a Berezinskii-Kosterlitz-Thouless (BKT) \cite{Berezinskii1971, Kosterlitz1973, KosterlitzJPhys74} transition with the Europium moments confined to the plane normal to the crystallographic $c$ axis by crystal anisotropy. This finding is based on two main observations: first, the variation of the specific heat $C$ in the vicinity of the phase transition agrees qualitatively and quantitatively with that of a BKT transition. 
In particular, the BKT scenario naturally explains the absence of a singularity at the transition point.
Furthermore, the anisotropic response of the specific heat to different field directions clearly points towards a strong ordering of the moments within the $\mathrm{Eu}$ planes. The reported findings are supported by numerical Monte-Carlo simulations of a classical anisotropic 2D Heisenberg spin system. 

Generally speaking, a BKT transition means that the state above the critical temperature can be viewed as a liquid of magnetic vortices and antivortices, while in the low temperature ordered phase only bound vortex-antivortex pairs are present. In a pure 2D case the average magnetic moment would thus be destroyed by spin-wave fluctuations also in the ordered phase. Weak interlayer coupling, as present in \ooff, promotes a small average in-plane moment formed at very large scales, while at smaller scale the behavior remains two-dimensional. This results in the fact, that the true phase transition in this system belongs to the universality class of the three-dimensional anisotropic Heisenberg model. However, these 3D effects are only relevant within a narrow range near the transition temperature and add fine features on the top of overall 2D behavior. Similar scenarios are realized in several layered magnetic compounds, such as K$_2$CuF$_4$\cite{HirakawaJPSJ82,HirakawaJAP1982} and Rb$_2$CrCl$_4$ \cite{CorneliusJPhys1986,BramwellJPSJ95}.

\begin{figure*}[tbh]
\includegraphics[width=0.98\textwidth]{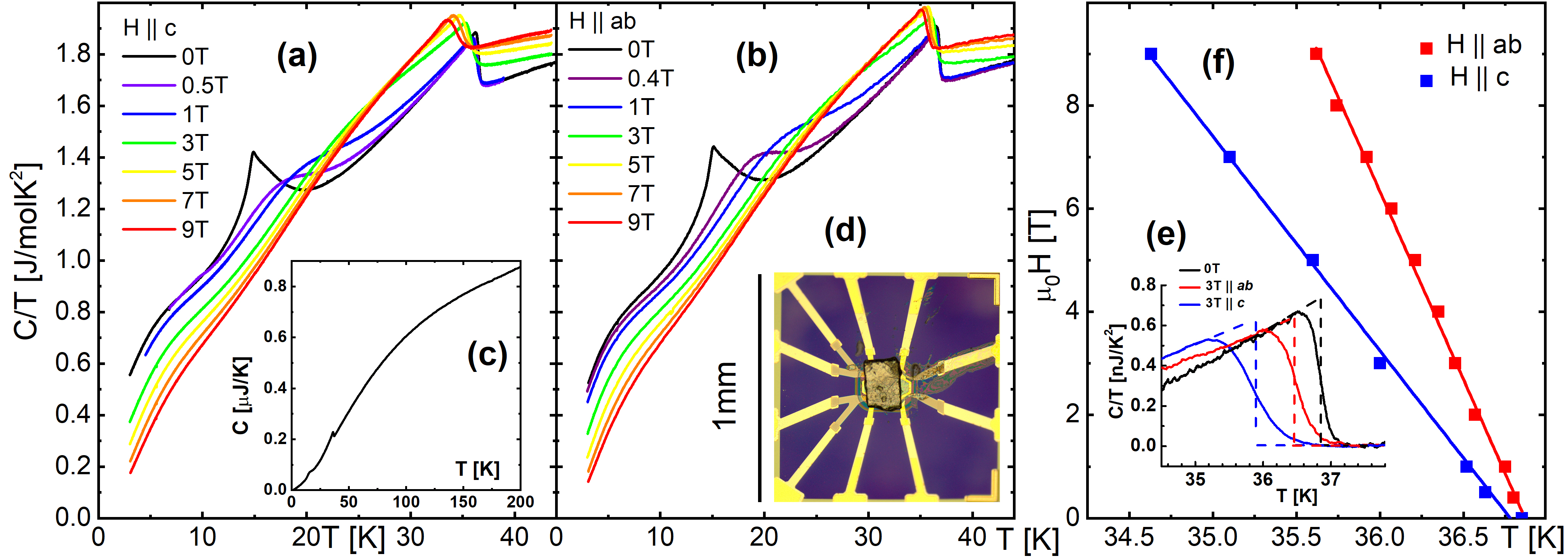}
\caption{
Entropy change ($C/T$) in single crystal {\ooff} and its dependence on the magnetic field strength when applied along \textbf{(a)} and perpendicular \textbf{(b)} to the crystallographic $c$ axis.
The superconducting transition at $T_{c} = 37\mathrm{K}$ and the magnetic transition at 15K are clearly visible in the zero-field calorimetric scan \textbf{(c)}, as obtained from a room-temperature cool down.
The microscope image \textbf{(d)} shows the $ac$ nanocalorimeter platform with a {\ooff} single crystal mounted at its center.
Following the evolution of the superconducting transition in an applied field, \textbf{(e)}, allows to extract the phase diagram \textbf{(f)} and to evaluate the superconducting anisotropy $\Gamma = 1.8$. The apparent discrepancy in the extrapolated $T_{c}$ is within the experimental uncertainty.
}
\label{fig:highfields}
\end{figure*}

\begin{figure*}[tbh]
\includegraphics[width=0.98\textwidth]{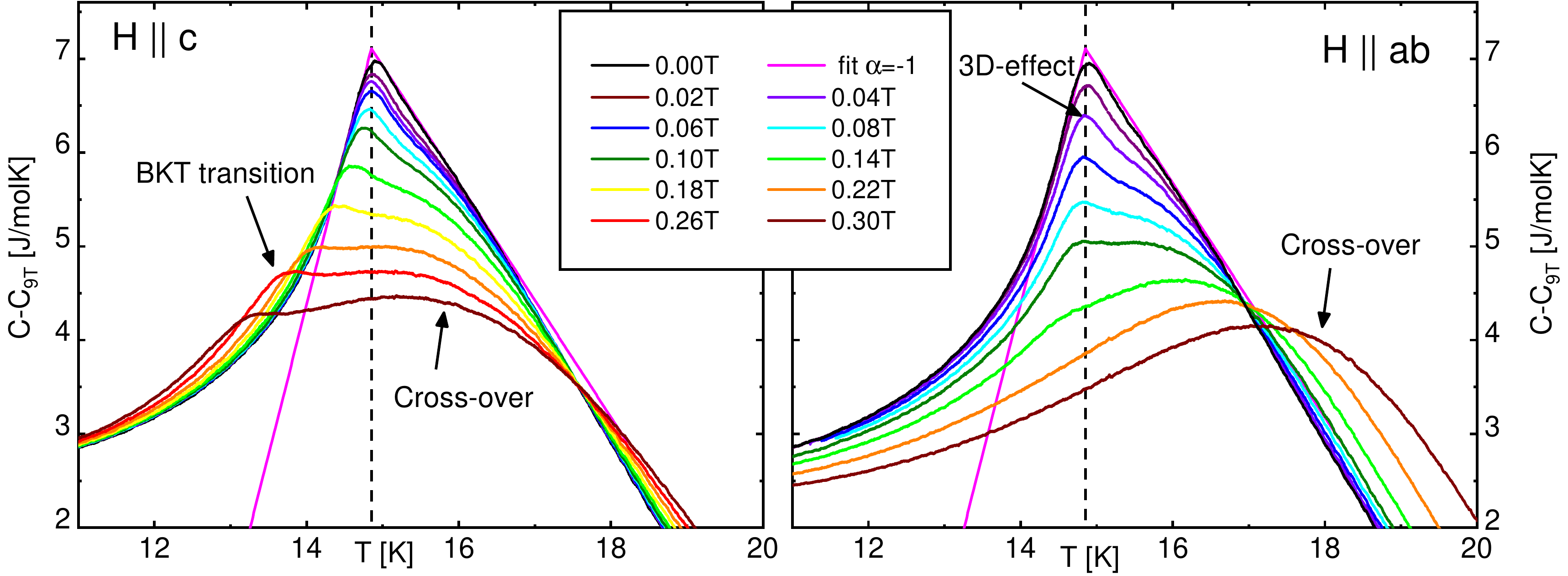}
\caption{specific heat subtracted by the 9T background curve around the magnetic transition upon applying fields from 0.02 to 0.3T  out-of-plane (left) and in-plane (right). The sharp kink indicating the ordering of the $\mathrm{Eu}^{2+}$ moments in the plane while the broad hump developing at higher temperatures shows the gradual magnetization of the sample parallel to the applied field.
}
\label{fig:smallfields}
\end{figure*}

The appearance of the superconducting phase below $T_{c} = 36.8(6)\mathrm{K}$ and a magnetic phase below $T_{m} = 14.9\mathrm{K}$ are clearly revealed in the calorimetric data obtained on zero-field cooling from room temperature down to $2\mathrm{K}$, see Fig.\ \ref{fig:highfields}. Whereas the superconducting transition temperature is extracted through an entropy conserving construction, see Fig.\ \ref{fig:highfields}(f), we determine the magnetic transition temperature from the position of the specific-heat cusp, which does not show signs of a first- or second-order phase transition. This observation is in line with previously reported results on polycrystalline $\mathrm{Cs} \mathrm{Eu} \mathrm{Fe}_{4} \mathrm{As}_{4}$ \cite{Liu2016a} and \ooff \cite{Liu2016b}, and should be contrasted to results on EuFe$_{2}$As$_{2}$ which show a singularity \cite{Jeevan2008b, Ren2008, Paramanik2014, Oleaga2014}. The variation of the specific heat in the vicinity of the phase transition, the specific heat can be expressed \cite{Wosnitza2007} as $C = a^{\pm}|t|^{-\alpha} + b(t)$. The first term captures the critical behavior near $t = 0$ with $t = T/T_{m}-1$ the reduced temperature, $a^{\pm}$ the critical amplitudes for $t<0$ ($-$) and $t>0$ ($+$), and $\alpha$ the critical exponent. The second term captures all regular contributions (e.g.\ from phonons) and is typically modeled by a linear form $b(t) = b_{0} + b_{1}t$ in a small temperature range around the transition. A non-divergent specific heat implies $\alpha < 0$, and hence the constant $b_{0} \equiv C(T_{m})$ assumes the value of the specific heat at the transition temperature. For each branch $t \lessgtr 0$, we find a critical exponent $\alpha \approx -1$; a highly unusual value. For the critical amplitudes we find $a^{+} = 18.5\mathrm{J/molK}$ and $a^{-} = 4.76\mathrm{J/molK}$ respectively, see fits in Fig.\ \ref{fig:smallfields}.

Contrary to earlier speculations \cite{Liu2016b}, we find that this transition with non-singular behavior is consistent with a Berezinskii-Kosterlitz-Thouless transition of the $\mathrm{Eu}^{2+}$ magnetic moments weakly influenced by 3D effects. A uniaxial anisotropy forces the moments to orient within the crystallographic $ab$ plane, effectively reducing the moment's degrees of freedom to that of a 2D $XY$ spin system. A more detailed justification shall be given below. The down-bending of the calorimetric data below $\sim 10\mathrm{K}$ is attributed to the quantum nature of the high spin $\mathrm{Eu}^{2+}$ moments \cite{Bouvier1991, Johnston2011,Smylie2018c}. In applied fields, the superconducting transition temperature is gradually suppressed; the effect is stronger, if the field is applied along the $c$ axis. The rate of $T_{c}$-suppression $dT_{c}/dH|_{ab} = 0.14\mathrm{K/T}$ and $dT_{c}/dH|_{c} = 0.25\mathrm{K/T}$, provides a uniaxial \emph{superconducting} anisotropy of $\Gamma \!=\! 1.8$, as shown in Fig.\ \ref{fig:highfields}. These values agree with complementary magnetization and transport measurements \cite{Smylie2018c, Stolyarov2018} on single crystal \ooff. No influence on the step height $\Delta C/T$ or the phase boundary from magnetism is detected in fields up to $9\mathrm{T}$.
In high fields, $0.4\mathrm{T} < \mu_{0} H < 9\mathrm{T}$, the cusp of the magnetic transition evolves into a broad magnetic hump with its center moving to higher temperatures. At the highest field (9T) these magnetic fluctuations extend up to about 100K---far above the superconducting transition---and provide a natural explanation for the reported negative, normal-state magneto-resistance\cite{Smylie2018c}. We attribute this hump to a field-induced polarization of the $\mathrm{Eu}^{2+}$ moments along the field direction and their associated fluctuations.

\begin{figure}[b!]
\includegraphics[width=0.98\linewidth]{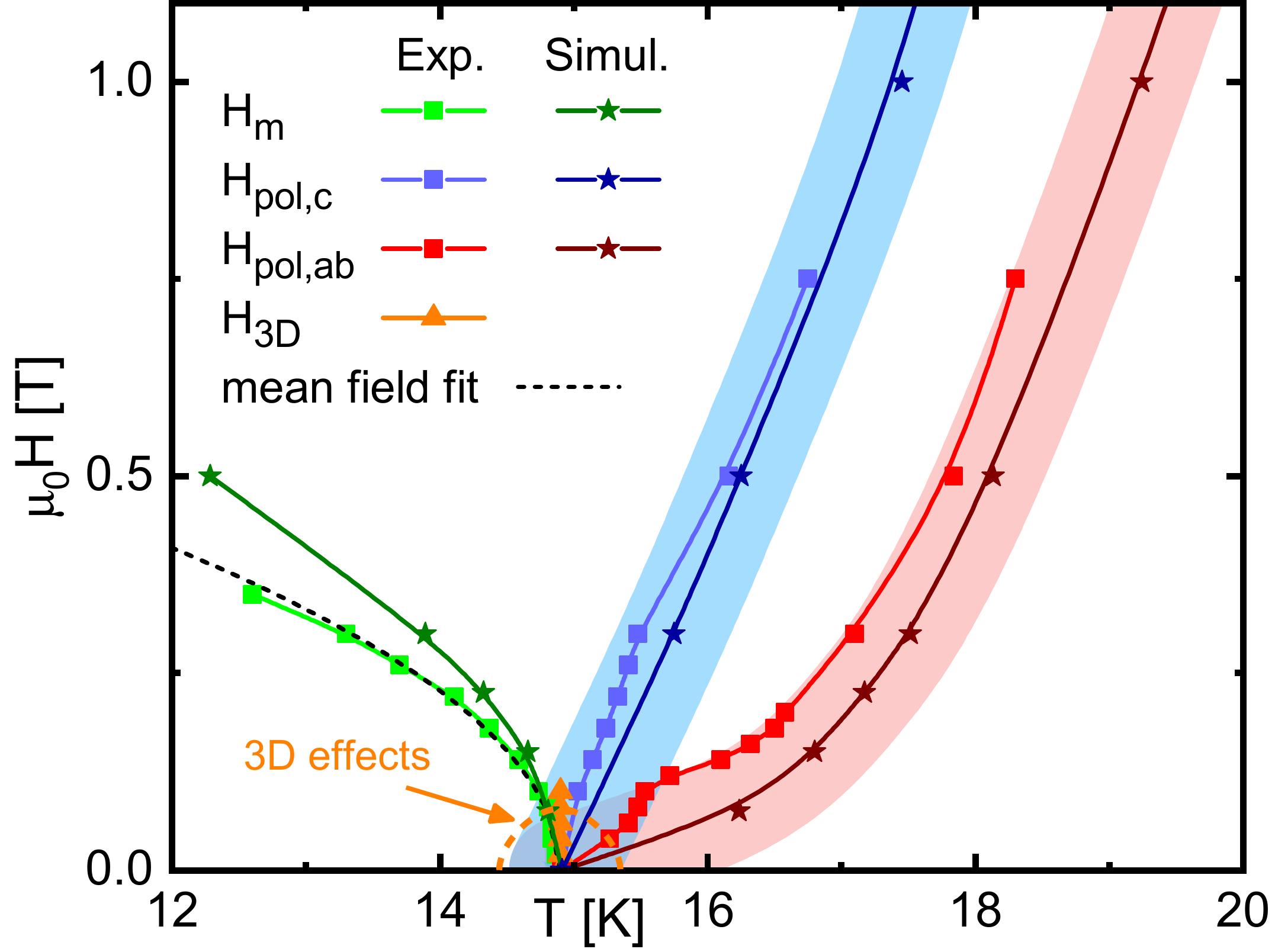}
\caption{Boundaries of the magnetic phase as obtained from measured and simulated calorimetric data. The transition to an ordered magnetic phase is shown by green symbols, and agrees well with the empiric law $H_m = H_0(1-T/T_m)^{1/2}$, when the field is applied along the $c$-axis. A broad hump in the specific heat marks the cross-over to a field-driven polarized state of $\mathrm{Eu}^{2+}$ moments and is shown for fields parallel (blue) and normal (red) to the $c$ axis.}
\label{fig:phaseboundary}
\end{figure}

For a more detailed analysis of the magnetic transition, we performed low-field calorimetric scans in the vicinity of $T_{m}$. Given the robust superconductivity (low $dT_{c}/dH$) and the clear separation of energy scales $\kB T_{m} \ll \kB T_{c}$, the (low-)field changes in the calorimetric data can be attributed to the magnetism. To accentuate these, we have to subtract an overall background. However, subtracting a phonon-type background turns out difficult because of other (in particular superconducting) contributions. We therefore subtract the $9\mathrm{T}$ specific heat data (field along $c$ axis). While the latter still contains magnetic and superconducting contributions, both are essentially featureless in the temperature range of interest, see Fig.\ \ref{fig:highfields}. 
%
As shown in Fig.\ \ref{fig:smallfields}, for small applied fields along the $c$ axis, the specific-heat cusp at the magnetic transition shifts to lower temperatures while broadening slightly and a shoulder in the specific heat appears on the high-temperature side. Defining the phase boundary $T_{m}(H)$ as the position of the cusp, see Fig.\ \ref{fig:phaseboundary}, a mean-field fit provides the empiric law $T_{m}(H) = T_{m} [1 - (H/H_{0})^{2}]$, with $\mu_{0} H_{0} \approx 0.93\mathrm{T}$. This suggests that at this field value the planar anisotropy is overcome at all temperatures, i.e. at zero temperature the magnetic moments fully align with the field normal to the $ab$ plane. A comparable saturation field can be deduced from low-temperature magnetization curves \cite{Smylie2018c}. For in-plane fields, the position of the cusp is almost field-independent, while its size is readily suppressed (disappearing at 0.14T) and a pronounced shoulder appears on the high-temperature side. As discussed below, we attribute the cusp to a weak 3D coupling between $\mathrm{Eu}$ layers.
The appearance of the high-temperature feature marks the onset of magnetic polarization, as discussed above. This hump is not a sharp phase boundary but should rather be understood as a crossover from a paramagnetically disordered to an ordered state. Due to anisotropy effects, this occurs more rapidly for in-plane than for out-of-plane fields.
\begin{figure*}[tbh]
\includegraphics[width=0.475\textwidth]{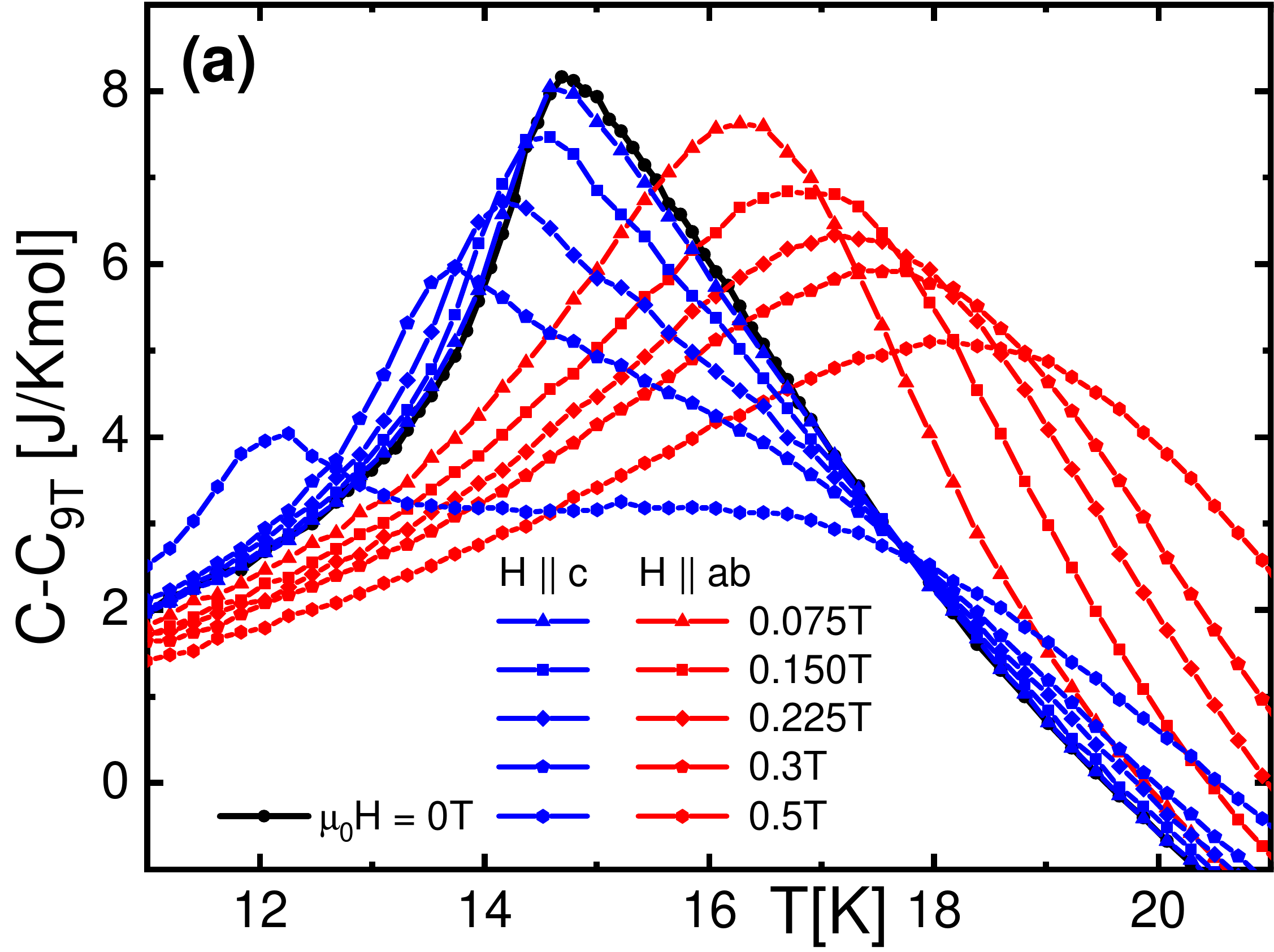}\quad
\includegraphics[width=0.476\textwidth]{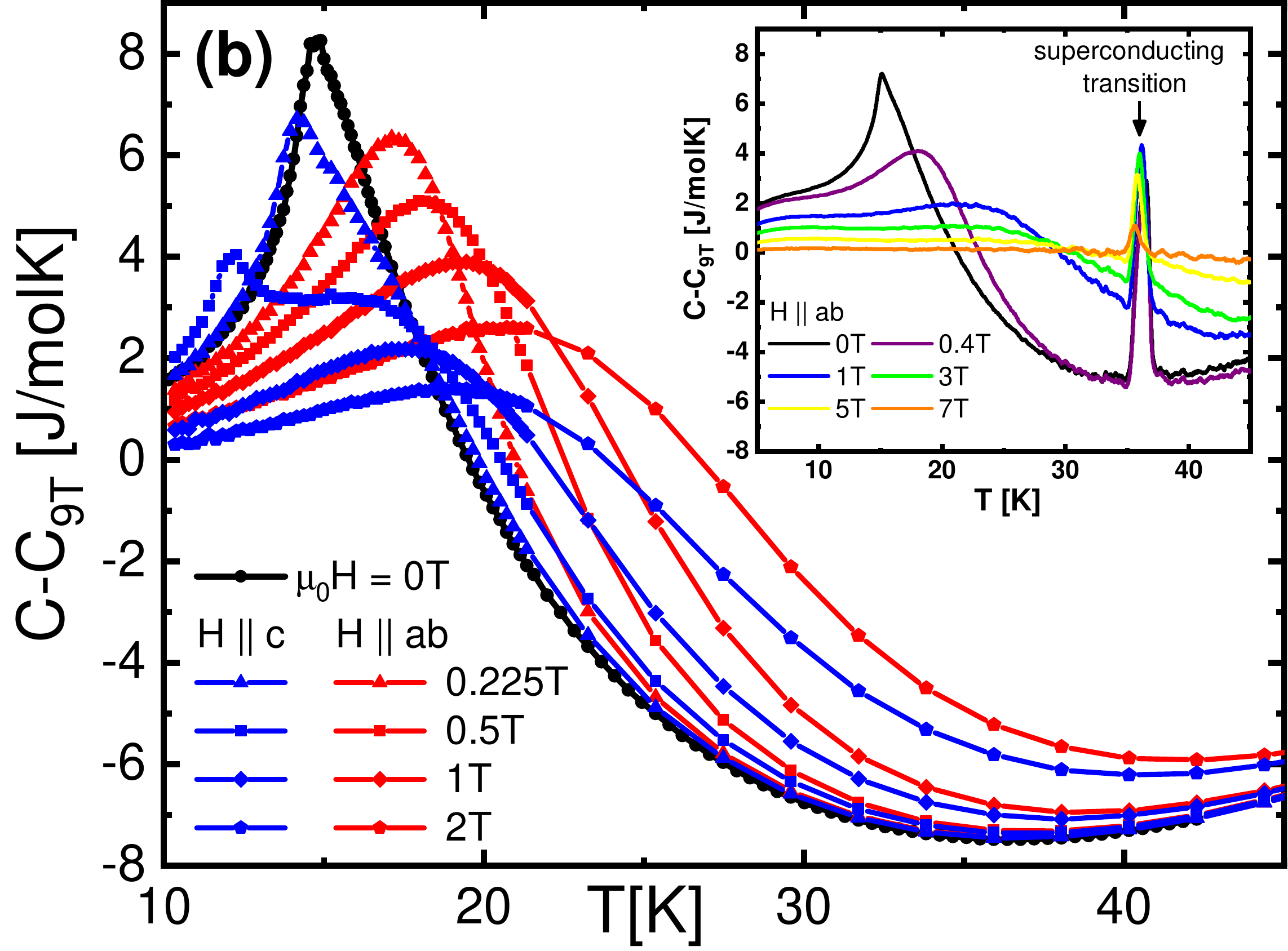}
\caption{Simulated specific heat of the anisotropic 2D Heisenberg spin system and its dependence on temperature for different magnetic fields and their orientations; red (in-plane), blue (out-of-plane), black (zero-field). For all curves the $9\mathrm{T}$-background data is subtracted [conversion to real units using \eq \eqref{eq:Hrealunits}]. \textbf{(a)} shows the low-field features and their anisotropic response near the magnetic transition. The specific heat at larger fields, and over a wider temperature range, is shown in \textbf{(b)}. The experimental signature of the superconducting transition near $37\mathrm{K}$, see inset, is not captured in the simulations.}
\label{fig:sim-high-low-field}
\end{figure*}

Further insight into the response of {\ooff} is gained through a detailed study of a model spin system describing the key features of this compound, implemented using a Monte-Carlo \cite{Metropolis1953,Hastings1970} algorithm, see Supplementary Material \ref{app:code}. More specifically, we have investigated the magnetic and thermodynamic properties of a two-dimensional square lattice of [Heisenberg-type, $\mathcal{O}(3)$] classical spins $\vec{s}_{i}$ governed by the Hamiltonian
\begin{align}\label{eq:anis2DHeisenberg}
   \mathcal{H} = &\ -J \sum_{\langle i,j\rangle} \vec{s}_{i}\vec{s}_{j}
               + K \sum_{i} (2 s_{i,z}^{2} -1)
               - \vec{h} \sum_{i} \vec{s}_{i}.
\end{align}
Here $J$ defines the isotropic coupling between nearest-neighbor spin pairs $\langle i,j \rangle$, $K$ introduces a uniaxial anisotropy in spin space. The last term describes the coupling to an external magnetic field $\vec{h}$. Without limiting the generality of the foregoing, we set $|\vec{s}_{i}| \!=\! 1$. A similar approach has been extensively used in the past to explore the 2D $XY$ model, see Refs.\ \sbonlinecite{Tobochnik1979, Janke1990, Gupta1992, Cuccoli1995, Sengupta2003}.

The simulated system is purely two-dimensional, and hence neglects the coupling between neighboring $\mathrm{Eu}$ layers. This choice is motivated by the observation that the parent non-superconducting compound $\mathrm{Eu}\mathrm{Fe}_{2}\mathrm{As}_{2}$ displays small interlayer interactions compared to the intralayer interactions. We expect the coupling between $\mathrm{Eu}$ layers to be even weaker in {\ooff}, as the separation between $\mathrm{Eu}$ layers doubled and two superconducting layers are in between. The interlayer coupling becomes relevant only at temperatures near the transition and for small magnetic fields.
In the Hamiltonian \eqref{eq:anis2DHeisenberg}, the spin anisotropy is modeled by a crystalline term $\propto s_{i,z}^{2}$. The $\mathrm{Eu}^{2+}$ ions have a vanishing angular momentum ($L = 0$), which excludes a crystalline anisotropy originating from spin-orbit coupling. However, the coupling between the $\mathrm{Eu}^{2+}$ moments and $\mathrm{Fe}$ $d$-electrons---the latter are known to feature an easy-plane anisotropy \cite{Wang2009,MeierPhysRevB2016}---naturally leads to such a term, see Supplementary Material \ref{app:crystal-anisotropy}. Other sources of anisotropy such as dipolar interactions, considered elsewhere\cite{Xiao2010}, are neglected. The anisotropic term causes the system to fall into the universality class of 2D $XY$ spin systems, where a BKT transition is known to occur at a finite temperature $T_{m} > 0$ \cite{HikamiProgrTheorPhys1980}. In contrast to our model, an isotropic (in spin space) 3D Heisenberg model with anisotropic nearest neighbor coupling ($J$ in-plane vs.\ $J' \!=\! \lambda J$ between $\mathrm{Eu}$ layers) fails to capture an anisotropic susceptibility, while the isotropic (in spin space) two-dimensional Heisenberg model does not undergo a phase transition at finite temperatures \cite{Polyakov1975, Brezin1976a, *Brezin1976b, Nelson1977, Shenker1980}. 

We investigate several response functions in this system: the (direction-dependent) magnetic susceptibility $\chi_{\alpha}(T)$ ($\alpha = x,y,z$), the specific heat $C(T, \vec{h})$, the total magnetization $\vec{S}(T,\vec{h}) = \sum_{i}\vec{s}_{i}$, and the spin-spin correlation function $G(r) \equiv \langle \vec{s}(0)\vec{s}(r)\rangle$. For convenience we introduce the temperature scale $T_{0} \equiv J/\kB$. From high-temperature simulations (typically $T/T_{0} \in [4,9]$), we fit the inverse magnetic susceptibility to a Curie-Weiss law $\chi_{\alpha}^{-1}(T) \propto T - \Theta_{C,\alpha}$ to extract the Curie temperatures $\Theta_{C,\alpha}$. A comparison between the measured and the simulated susceptibility can be found in the Supplementary Material \ref{app:Magn}. Any non-zero value of $K$ results in an anisotropy between the in-plane ($\Theta_{C,x}$) and out-of-plane ($\Theta_{C,z}$) Curie temperature. By comparing the anisotropy ratio $\Theta_{C,x}/\Theta_{C,z}$ with the reported \cite{Smylie2018c} value $1.075$ for {\ooff} obtained from magnetization measurements, we find an agreement for the specific value $K = 0.1 J$, where $\Theta_{C,x} = 1.20T_{0}$ and $\Theta_{C,z} = 1.12T_{0}$. All further simulations are performed for this anisotropy parameter. The influence of the anisotropy parameter on the shape of $C(T)$ dependence at $h=0$ and the location of the cusp is considered in the Supplementary Material \ref{app:Anis}.

In zero magnetic field, the simulated specific heat shows a clear cusp at $T_{m}/T_{0} = 0.7$, a value that we identify with the transition temperature $T_{m} = 14.9\mathrm{K}$ of the calorimetric experiment. It is known however, that the true BKT transition temperature $T_{\sss \BKT}$ is slightly below the specific heat cusp. The correlation function is expected to decay as a power law $r^{-1/4}$ at the transition, providing a value $T_{\sss \BKT} \!=\! 0.66 T_{0}$, i.e., about $6\%$ below the cusp in the specific heat. At the same time, the correlation function decays as $G(r) \propto \exp[-r/\zeta(T)]$, with a correlation lengths $\ln[\zeta(T)] \propto (T-T_{\sss \BKT})^{-1/2}$ that diverges upon approaching the transition from above. Evaluation of $\zeta(T)$ and its singular behavior yields a consistent result, see Supplementary Material \ref{app:CorrFun}.

At finite fields, the calorimetric and magnetic responses strongly depend on the field orientation. For fields applied along the spin plane, the $U(1)$ circular degeneracy is lifted and no BKT transition occurs. The system's response follows a typical ferromagnetic behavior (gradual magnetization upon cooling) reaching a fully ordered state at lowest temperatures. The specific heat gradually broadens and shifts to higher temperatures. On the contrary, a field applied perpendicular to the spin lattice preserves the $U(1)$ rotational symmetry and the BKT transition shifts to lower temperatures. Here the magnetic field acts as an anisotropic term favoring the spin orientation along the $z$ axis, hence retarding the transition to an in-plane spin orientation. The numerical simulations are in excellent qualitative and quantitative agreement with the experimental data, see Figs.\ \ref{fig:smallfields} and \ref{fig:sim-high-low-field}. Additionally, the simulations reproduce the behavior of the magnetization and the susceptibility which is discussed in the Supplementary Material \ref{app:Magn}.
The phase boundaries extracted from the simulation data (converted to appropriate units) are shown in Fig.\ \ref{fig:phaseboundary}. The green curve corresponds to the suppression of the BKT transition due to a field normal to the spin plane. The two other curves correspond to a crossover where magnetic moments are polarized along the field (blue, $H \| c$ and red, $H \| ab$).
The simulation result reproduces the experimental data extremely well, with only minor deviations for low fields along the $ab$ plane. This difference is attributed to 3D-effects close to the transition, that are not accounted for in the simulations.

Having identified realistic values for the dimensionless parameters $T/T_{0}$ (from calorimetry) and $K/J$ (from high-temperature magnetization), we can rewrite the model Hamiltonian in \eq \eqref{eq:anis2DHeisenberg} in a dimensional form
\begin{align}\label{eq:Hrealunits}
   \mathcal{H}_{\mathrm{real}} = &\ -\mathcal{J} \sum_{\langle i,j\rangle} \vec{m}_{i}\vec{m}_{j}
               + 2\mathcal{K} \sum_{i} m_{i,z}^{2}
               - \vec{H} \vec{M},
\end{align}
where a constant shift has been omitted. Here $\vec{M} = \sum_{i} \vec{m}_{i}$ denotes the total magnetization of the individual constituents $|\vec{m}_{i}| \approx 7\muB$, $\mathcal{J} = 0.6\times10^{-23}\mathrm{J}/\muB^{2}\ (=10\mathcal{K})$ providing the relevant energy scale for the ferromagnetic interactions (the anisotropy). It is useful to express the simulated fields $h$ in dimensional units via $h \to \mu_{0} H = 4.53 h [\mathrm{T}]$.

The numerical results are backed up by a high-temperature expansion of the model described by Eq.\ \eqref{eq:anis2DHeisenberg}, see Supplementary Material \ref{sec:high-T-exp}. Here the anisotropy ratio in the Curie temperature takes the simple form
\begin{align}\label{}
   \Theta_{C,x}/ \Theta_{C,z} = (1+K/5J) / (1-2K/5J)
\end{align}
and yields the value 1.06 for $K = 0.1J$. This relation reiterates that for an easy-plane anisotropy $K > 0$ the ratio of Curie temperatures $\Theta_{C,x}/\Theta_{C,z}$ is larger than unity, whereas an easy-axis system ($K < 0$) has $\Theta_{C,x}/\Theta_{C,z} < 1$. Note that the sign of the anisotropy may change for different $\mathrm{Eu}$-containing compounds. We find that the presumed 'high-temperature' range $T/T_{0} \in [4, 9]$ (corresponding to $50 \mathrm{-}200\mathrm{K}$) is only captured properly when the high-temperature expansion is taken to quartic order in $\beta\mathcal{H}$ [the susceptibility is expanded to cubic order in $\beta = (\kB T)^{-1}$]. This explains the noticeable discrepancy between the 'exact' values $\Theta_{C,x}/T_{0} = 4/3 + 4K/15J\ (= 1.36)$ and $\Theta_{C,z}/T_{0} = 4/3 -8K/15J\ (= 1.28)$ obtained in the high-temperature limit and their numerical counterparts 1.20 and 1.12, see above.

We have assumed that the third dimension, perpendicular to the easy-plane, plays a marginal role in the calorimetric response of the magnetic order. A weak coupling $J' = \lambda J$ ($|\lambda| \ll 1$) between ferromagnetically ordered $\mathrm{Eu}$ layers will add a fine structure on top of the leading features. Very close to the transition, when the correlation length $\zeta(T)$ reaches the in-plane length scale $1/\sqrt{\lambda}$ at the temperature\cite{HikamiProgrTheorPhys1980}
$T-T_{m} \sim T_{m}\ln^{-2}(1/\lambda)$,
the three-dimensional effects lead to full ordering of the system. On general ground, these effects should sharpen the specific heat cusp in close vicinity of the transition \cite{Janke1990, Sengupta2003}.
The nature of this three-dimensional order depends on the interlayer interactions: while a simple coupling $J'$ between neighboring layers results in a trivial ferro- ($J'>0$) or A-type antiferromagnet ($J' < 0$), more complicated helical, and fan-like orders can be found if longer-range interactions along $z$ are considered\cite{Nagamiya1962, Johnston2017}. In the latter cases there is the typical in-plane magnetic field scale $B_{\mathrm{3D}}=J'/|\vec{m}|$ aligning the moments in different layers in the same directions and eliminating the magnetic transition.

In conclusion, we have investigated the magnetic transition in {\ooff} by specific heat measurements and by Monte-Carlo simulations. The magnetic transition at 14.9K shifts to lower temperatures in fields along the $c$ axis. This is well reproduced by the simulations of the 2D anisotropic Heisenberg system. This allows us to identify the $ab$ plane as the magnetic easy plane and the specific-heat curve indeed shows a dominant BKT character. A magnetic field normal to the $\mathrm{Eu}$ layers shifts the magnetic transition to lower temperature. Applying the field along the $\mathrm{Eu}$ planes lifts the rotational symmetry required for a BKT transition. The latter is replaced by a broad crossover from a paramagnetically disordered to an field-ordered state. With a quantitative comparison between our simulation and experimental data, we have extracted the coupling constants $\mathcal{J} = 0.6\times10^{-23}\mathrm{J}/\muB^{2}$ and the anisotropy $\mathcal{K} = 0.1 \mathcal{J}$. The extraction of the phase boundary of the BKT transition and the crossover lines for in- and out-of-plane fields further underline the excellent agreement between experiment and simulations. 

The unique feature of {\ooff} is that the magnetic transition takes place deep inside the superconducting phase. We expect that the superconductivity has almost no influence on the intralayer exchange interaction between Eu moments and may only modify the interlayer interactions. Therefore, the direct impact of superconductivity on magnetism is likely to be minor. The effects caused by the opposite influence of magnetism on superconductivity are expected to be more pronounced. The presence of the magnetic subsystem with a large susceptibility drastically modifies the macroscopic magnetic response of the superconducting material in the mixed state \cite{Vlasko:ArXiv2019}. 
The source of the microscopic interaction between the  magnetic and superconducting subsystems is an exchange coupling between the Eu moments and Cooper pairs. Even though this coupling is not strong enough to completely destroy superconductivity, like, e.g., in ErRh$_4$B$_4$ \cite{ErRh4B4}, it may cause a noticeable suppression of superconducting parameters at the magnetic transition. 
Having established the nature of the robust magnetic order, this work serves as a starting point for further exploring the 
phenomena related to the influence
of magnetism on the superconducting state.

\begin{acknowledgments}
We are thankful for the helpful discussions with Matthew Smylie and Andreas Rydh.
This work was supported by the U.S. Department of Energy, Office of Science, Basic Energy Sciences, Materials Sciences and Engineering Division. K.\,W. and R.W. acknowledge support from the Swiss National Science Foundation through an Early Postdoc Mobility fellowship.
\end{acknowledgments}
\vfill

\pagebreak
\clearpage
\onecolumngrid
\appendix

\begin{center}
\Large \textbf{Supplementary Material}
\end{center}
\renewcommand\appendixname{}

\section{Experiment}

For our calorimetry experiment, we mounted a small, platelet-shaped {\ooff} single crystal, grown in RbAs flux \cite{Bao2018}, onto a nanocalorimeter platform \cite{Tagliati2012, Willa2017} using Apiezon grease, see Fig.\ \ref{fig:highfields}. The probe was then inserted into a three axis vector magnet (1T-1T-9T), where the field axes were aligned with the crystal axes within $\pm$3 degrees. The specific heat data, as obtained from $ac$ measurements ($f = 1\mathrm{Hz}$ and  $\delta T \sim 0.1\mathrm{K}$), was recorded with a Synktex lock-in amplifier.

\section{Numerical routine}\label{app:code}
Given a spin configuration, a Monte-Carlo iteration step consists in evaluating the energy change $\delta E$ induced by virtually substituting an existing spin $\vec{s}_{i}$ at site $i$ by a new spin $\tilde{\vec{s}}_{i}$. If $\delta E$ is negative, the replacement $\vec{s}_{i} \to \tilde{\vec{s}}_{i}$ is effected. In the opposite case $\delta E > 0$, the replacement is performed with a reduced probability $p = \exp(- \delta E/T)$. Numerically, this evaluation and update procedure is particularly suited for local Hamiltonians, where each spin only interacts with few neighboring spins.
Repeating this step for each spin of the lattice then defines one \emph{pass}. Thermal properties such as the system's average energy, magnetization, and their respective fluctuations can then be studied by evolving the system through many passes. In general, the new spin $\tilde{\vec{s}}_{i}$ is chosen by picking a random unit vector. At low temperatures, however, when the rate of accepted spin changes drops below a certain threshold (typically 0.5), we reduce the explored phase space to a cone centered around $\vec{s}_{i}$ and with an opening angle $\Omega$, i.e. $\arccos(\vec{s}_{i} \tilde{\vec{s}}_{i}) \leq \Omega$. The value of $\Omega \leq \pi$ is adjusted to yield an acceptance rate close to the threshold acceptance rate.

In our implementation, we simulate a square lattice of $L\times L$ spins (typically $L = 100$); initialized in a random spin configuration (the same for each run). Given a fixed parameter set $(T, \vec{h})$ (all simulations shown in the main text are obtained with $J = 10K = 1$), we run the simulation through $N_{p}/2$ passes for thermalization after which observable quantities and their fluctuations are evaluated over the next $N_{p}$ passes. The spins are visited typically $N_{p} = 10^{7}$ times in a random order (reshuffled after each pass). The total energy $E$, its square, the total magnetization $\vec{S} = \sum_{i}\vec{s}_{i}$, and its component's square are (time-)averaged over $10^{4}\text{-}10^{5}$ passes and written to file for post-processing.

The magnetic susceptibility and the specific heat (both per spin) are derived from the above quantities via
\begin{align}\label{}
   \chi_{\alpha} = \frac{1}{L^{2}} \frac{\langle S_{\alpha}^{2} \rangle - \langle S_{\alpha} \rangle^{2}}{\kB T}
   \text{ \ and \ }
   C = \frac{\kB}{L^{2}}\frac{\langle E^{2} \rangle - \langle E \rangle^{2}}{(\kB T)^{2}}
\end{align}
respectively. An equivalent expression for the specific heat be $L^{-2}\partial \langle E\rangle / \partial T$ was used to check the accuracy of the results.

\section{Properties the anisotropic 2D Heisenberg model}

\subsection{High-temperature expansion}\label{sec:high-T-exp}
When writing the Hamiltonian \eqref{eq:anis2DHeisenberg} using the spin representation $\vec{s}_{i} = (\cpi\sti,\spi\sti,\cti)$ we find
\begin{align}\label{}
	\mathcal{H} = \!-J\!\sum_{\langle i,j\rangle} [\sti\stj \cos(\phi_{i}\!-\!\phi_{j}) + \cti\ctj]
	+ \!K\!\sum_{i} \cos(2\theta_{i})
	-\! \sum_{i} [h_{x}\cpi\sti + h_{y}\spi\sti
	+ h_{z}\cti].
\end{align}
The system's partition function reads $Z = \llangle \exp(-\beta \mathcal{H}) \rrangle$
with $\beta = (\kB T)^{-1}$ and $\llangle \dots \rrangle = \prod_{i} (4\pi)^{-1} \int \dots \sti d\theta_{i}d\phi_{i}$ the uniform average over the unit sphere. Observable quantities can be evaluated from the free energy per spin $F = \ln(Z)/\beta N$ , which (at high temperatures) can be approximated by
\begin{align}\label{}
	F \approx F^{{(4)}} \equiv\frac{-1}{\beta N} \Big[
	&- \beta \llangle \mathcal{H} \rrangle
	+ (\beta^{2}/2) \big(\llangle \mathcal{H}^{2} \rrangle - \llangle \mathcal{H} \rrangle^{2}\big)
	- (\beta^{3}/6) \big(\llangle \mathcal{H}^{3} \rrangle - 3 \llangle \mathcal{H}^{2} \rrangle \llangle \mathcal{H} \rrangle + 2 \llangle \mathcal{H} \rrangle^{3}\big) \nonumber\\
	&+ (\beta^{4}/24) \big(\llangle \mathcal{H}^{4} \rrangle - 4 \llangle \mathcal{H}^{3} \rrangle \llangle \mathcal{H} \rrangle - 3 \llangle \mathcal{H}^{2} \rrangle^2 + 12 \llangle \mathcal{H} \rrangle^2 \llangle \mathcal{H}^{2} \rrangle -6 \llangle \mathcal{H} \rrangle^4\big) \Big]
\end{align}
to fourth order in $\mathcal{H}$, where $\llangle \mathcal{H}^{n} \rrangle$ is the $n$-th moment of the Hamiltonian. Going through the calculation of each moment, one finally arrives at the expression
\begin{align}\label{}
	F^{{(4)}} &= - K/3  - \beta \Big[\frac{J^{2}}{3} + \frac{8 K^{2}}{45} + \frac{h^{2}}{6}\Big] - \beta^{2} \Big[\frac{2 J h^{2}}{9} - \frac{64 K^{3}}{2835} + \frac{2 K (h_{x}^{2} + h_{y}^{2} - 2 h_{z}^{2})}{45}\Big] \\\nonumber
	&\quad - \beta^{3} \Big[\frac{7 J^{4}}{270} - \frac{64 K^{4}}{14175} - \frac{h^{4}}{180} + \frac{32 J^{2}K^{2}}{675} + \frac{2 J^{2} h^{2}}{9} + \frac{112 J K - 8 K^{2}}{945} (h_{x}^{2} + h_{y}^{2} - 2h_{z}^{2})\Big]
\end{align}
Whereas up to linear order in $\beta$ the anisotropic term $\propto K$ merely shifts the energy, the spin response becomes truly anisotropic with the term $\propto \beta^{2} K h_{\alpha}^{2}$. The specific heat per spin $C = - \kB  \beta^{2} d U/ d \beta$ is obtained from the average energy per spin $U = -(1/N) d \ln Z/d\beta = d(\beta F)/d\beta$ and amounts to
\begin{align}\label{}
	C &= \kB  \beta^{2}
	\Bigg\{ 
	\Big( \frac{2J^{2}}{3} + \frac{16 K^{2}}{45}\Big)
	- \beta \frac{128 K^{3}}{945}
	+ \beta^{2} \Big( \frac{14 J^{4}}{45} + \frac{128 J^{2} K^{2}}{225} - \frac{256 K^{4}}{4725} \Big) \nonumber\\
	&\qquad + \frac{h^{2}}{3}
	+ \beta \Big[ \frac{4 J h^{2}}{3} + \frac{4 K}{15}(h_{x}^{2} + h_{y}^{2} - 2h_{z}^{2}) \Big]
	+ \beta^{2} \Big[ \frac{8 J^{2} h^{2}}{3} + \frac{32(14J - K)K}{315}(h_{x}^{2} + h_{y}^{2} - 2h_{z}^{2}) -\frac{h^{4}}{15}\Big]
	\Bigg\}.
\end{align}
Here the first line corresponds to the zero-field limit and the second line includes all field terms. Components of the average spin are obtained from $s_{\alpha} = (1/\beta N) d \ln Z/ d h_{\alpha} = -dF/dh_{\alpha}$, and the susceptibilities $\chi_{\alpha} = \lim_{h\to0} [s_{\alpha} / h_{\alpha}]$ read
\begin{align}\label{}
	\chi_{x} = \chi_{y} &= 
	\frac{\beta}{3}
	+ \beta^{2} \Big( \frac{4J}{9} + \frac{4K}{45}\Big)
	+ \beta^{3} \Big( \frac{4J^{2}}{9} + \frac{32 J K}{135} - \frac{16 K^{2}}{945}\Big)\\
	\chi_{z} &= \frac{\beta}{3}
	+ \beta^{2} \Big( \frac{4J}{9} - \frac{8K}{45}\Big)
	+ \beta^{3} \Big( \frac{4J^{2}}{9} - \frac{64 J K}{135} + \frac{32 K^{2}}{945}\Big)
\end{align}
For high temperatures $J\beta \ll 1$, the susceptibility follows a Curie-Weiss law $\chi_{\alpha} \approx \chi_{0}^{\alpha} /(T/\Theta_{C, \alpha} - 1)$, with $\Theta_{C, \alpha}$ the orientation-dependent Curie-Weiss temperature, and $\chi_{0}^{\alpha}$ a constant. We have
\begin{align}\label{}
	\chi_{0}^{x} 
	&= \chi_{0}^{y} = 5 / [4 (5 J + K)] & \chi_{0}^{z} &= 5 / [4 (5 J - 2K)]\\
	\kB  \Theta_{C,x} &= \kB  \Theta_{C,y} = (4/15)(5 J + K) & \kB  \Theta_{C,z} &= (4/15)(5 J - 2K)
\end{align}
When measuring the inverse susceptibility, the curves along $x$ and $z$ are shifted by a constant, whereas the slopes $1/\chi_{0}^{\alpha}\Theta_{C,\alpha}$ are the same. The ratio of the two Curie temperatures (in-plane vs. out-of-plane) then reads $\Theta_{C,x} / \Theta_{C,z} = (5 + K/J)/(5 - 2 K/J)$, while the sign of the Curie-Weiss temperature indicates whether the spin coupling is dominantly ferromagnetic ($J>0$) or antiferromagnetic ($J<0$), the ratio $\Theta_{C,x} / \Theta_{C,z}$ is informative about the nature of the anisotropy. In fact the system is an easy-plane ($XY$-type) magnet if $\Theta_{C,x} / \Theta_{C,z} > 1$. If $\Theta_{C,x} / \Theta_{C,z} < 1$ the magnet has easy axis (Ising-type) with moments orienting preferentially normal to the spin plane.
\begin{figure}[tbh]
	\centering
	\includegraphics[width=0.96\textwidth]{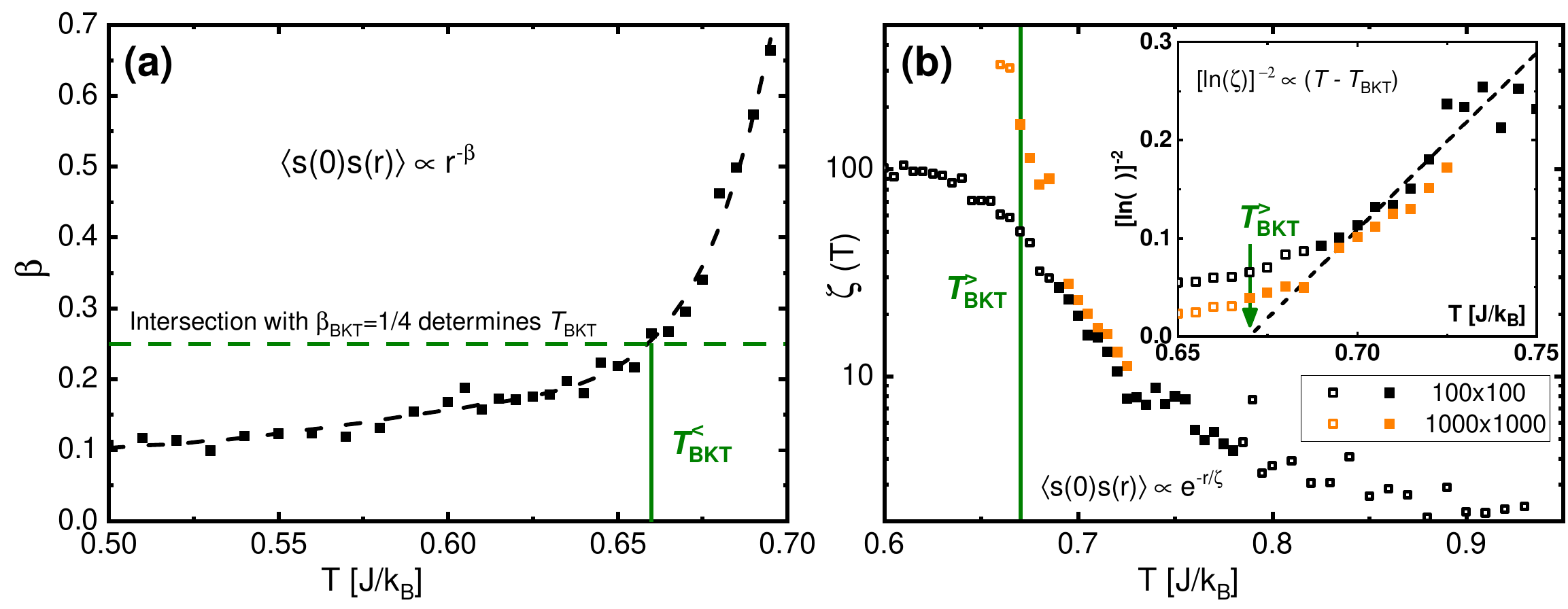}

	\caption{An accurate definition of the BKT transition temperature involves studying the temperature-dependence of the correlation function $G(r) = \langle \vec{s}(0) \vec{s}(r) \rangle$. \textbf{(a)} Below $T_{\BKT}$, the latter decays as a power law $G(r)\propto r^{-\beta}$ with $\beta_{\BKT} = 1/4$. The intersection of a fit (black dashed) with $1/4$ provides $\kB T_{\BKT}^{<}/J = 0.66$. \textbf{(b)} Above the transition, the correlation length decays exponentially, $G(r)\propto e^{-r/\zeta}$, with the correlation length $\ln \zeta(T) \propto (T-T_{\BKT})^{-1/2}$. The linear fit  $[\ln(\zeta)]^{-2} = -2.41 + 3.60 T/T_0$ (black dashed) to the black and orange filled symbols provides $T_{\BKT}^{>} \approx 0.67 T_0$. This extraction is more sensitive to the system's finite size, as it requires to approach very close to the transition, where the correlation length diverges. Shown here is the correlation length for the typical system ($100\times100$ spins, black) and selected points have been computed for a large system ($1000\times1000$ spins, orange).}
	\label{Fig:PowerCorrLength}
\end{figure}
\subsection{Behavior of the correlation functions and the exact location of the BKT temperature}\label{app:CorrFun}
As mentioned in the main text, the cusp in specific heat just corresponds to an approximate location of the transition. In contrast to conventional magnetic transitions, the long-range order does not emerge below $T_{\BKT}$. In order to extract the true BKT transition temperature, we have to analyze the moment correlation function $G(\vec{r})=\langle \vec{s}(0) \vec{s}(r) \rangle$.  The long-range behavior of $G(\vec{r})$ changes qualitatively at $T_{\BKT}$, it decays exponentially in the paramagnetic state $G(\vec{r}) \propto \exp(-r/\zeta)$ and below $T_{\BKT}$ the decay is algebraic $G(\vec{r}) \propto r^{-\beta}$. Moreover, as follows from the theory of the BKT transition\cite{KosterlitzJPhys74},   the moment correlation length diverges as $\zeta(T) \propto \exp\!\big[b\sqrt{T_{\BKT}/(T\!-\!T_{\BKT})}\big]$ for $T\rightarrow T_{\BKT}+0$ and the value of the power exponent $\beta$ at  $T_{\BKT}$ is exactly $1/4$.
Figure \ref{Fig:PowerCorrLength} shows temperature dependence of the power exponent $\beta$ and the correlation length $\zeta$ extracted from fits of the numerically computed correlation function. Extracting the true transition temperatures below ($T_{\BKT}^{<} = 0.66 T_{0}$) and above ($T_{\BKT}^{>} \approx 0.67 T_{0}$) leads us to conclude, that the transition is at $T_{\BKT}=0.67 T_{0}$ which is 5\% below the cusp position seen at $T_m=0.7 T_{0}$.

\subsection{Behavior of susceptibility and magnetization}\label{app:Magn}

\begin{figure}[t!]
	\includegraphics[width=0.96\textwidth]{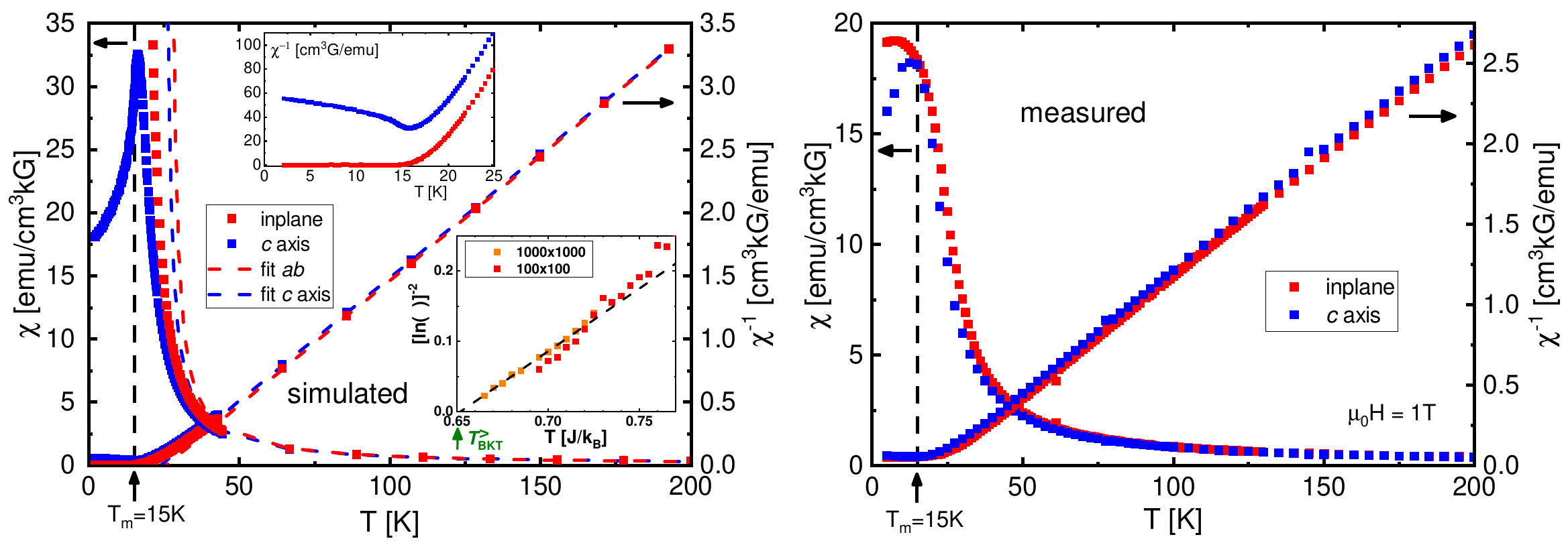}
	\caption{Comparison of the simulated linear susceptibilities (\emph{left}) and the measured susceptibilities at fixed magnetic field $\mu_{0} H = 1$T (\emph{right}, published earlier in Ref.\ \sbonlinecite{Smylie2018c}). The simulated susceptibility has been converted to real units by using the density of $0.5\times 10^{22}$ Europium atoms per $\mathrm{cm}^{3}$.  The dashed lines in the simulation plot show fits to the Curie-Weiss dependence discussed in the main text and the higher inset highlights the susceptibility's low-temperature dependence. The lower inset verifies  the BKT scaling $[\ln(\chi(T))]^{-2} = -1.16 + 1.78 T/T_0$ expected near the transition giving a transition at $T_{\BKT}^{>} \approx 0.65 T_{0}$ close to the value extracted from the correlation length.}
	\label{fig:SimSus}
\end{figure}

\begin{figure}[tbh]
\includegraphics[width=0.96\textwidth]{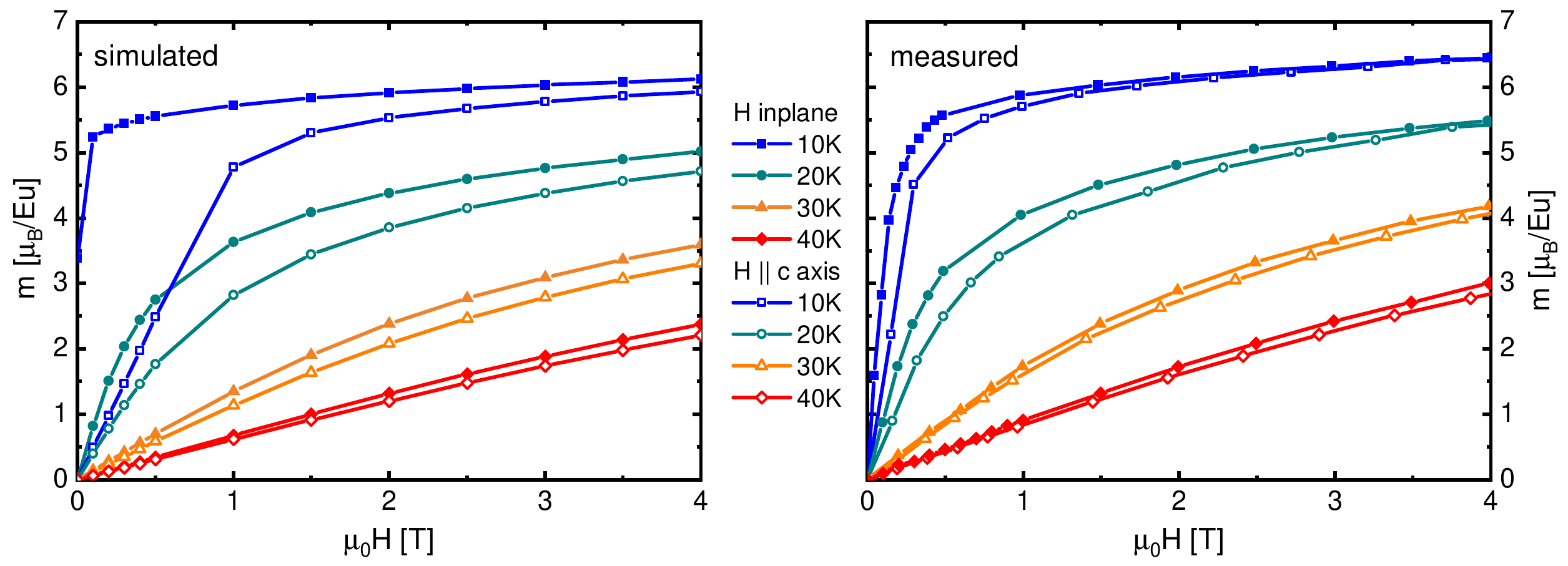}
\caption{Simulated (left) and measured (right) isothermal magnetization curves at $T = 10, 20, 30,\, \text{and } 40\mathrm{K}$, for fields applied in and out of the basal plane. The magnetization is normalized to the magnetization per $\mathrm{Eu}$ atom. The measured magnetization data as reprinted from Ref. \sbonlinecite{Smylie2018c} at temperatures below the superconducting transition were calculated from measured magnetization hysteresis curves by evaluating the symmetric part which in a first approximation removes the effects due to vortex pinning as described there.}
\label{fig:SimMagn}
\end{figure}

We have investigated to some detail the magnetic response of the spin system for the anisotropy $K = 0.1 J$ (discussed in the main text). Figure \ref{fig:SimSus} compares the temperature dependencies of the computed linear susceptibilities and the measured susceptibilities at fixed magnetic field\cite{Smylie2018c} $\mu_{0} H = 1$T.
The comparison is only justified above 30 K, because, as evident from Fig.\ \ref{fig:SimMagn}, the susceptibility at lower temperatures becomes nonlinear at this field. Furthermore, the behavior of the experimental susceptibility data at smaller fields is strongly influenced by the appearance of superconductivity. In the high-temperature range the model describes well the temperature dependencies of susceptibilities and their anisotropy. A linear fit to the high-temperature part of the measured inverse susceptibility leads to the Curie temperatures for the in- and out-of-plane directions. This information was extracted and used to determine the anisotropy parameter for the simulations.

In the vicinity of the BKT transition, the linear susceptibility diverges as $\chi(T)\propto \zeta^2 \propto \exp\!\big[2b\sqrt{T_{\BKT}/(T\!-\!T_{\BKT})}\big]$. The simulated susceptibility indeed shows this behavior, as illustrated in 
the lower inset of Fig.\ \ref{fig:SimSus} (left), where we plot the temperature dependence of $[\ln (\chi)]^{-2}$ for two system sizes. The linearity of this dependence indicates the validity of the BKT scaling. The slope should differ from the slope of the correlation length by a factor of 2 which is in very good agreement (1.78 vs. 3.59).

Figure \ref{fig:SimMagn} presents  the simulated and measured isothermal magnetization curves. The experimental magnetization data at temperatures below the superconducting transition were calculated from magnetization hysteresis curves by evaluating the symmetric part which in a first approximation removes the effects due to vortex pinning, as described in Ref.\ \sbonlinecite{Smylie2018c}.
The isothermal magnetization shows an increasing anisotropy upon decreasing the temperature. Furthermore, the saturation field, i.e. the field where the magnetization reaches (almost) full polarization rapidly decreases with decreasing temperature. Below the magnetic transition, the simulation's in-plane magnetization is non-zero even at zero magnetic field. This is an artifact of the system's finite size; For its average to vanish---as expected for a BKT phase---an exponentially large simulation time is required. Overall, we find a very good agreement between the simulated and measured magnetization.

\subsection{The role of magnetic anisotropy: From Heisenberg to 2D $XY$ model}\label{app:Anis}
\begin{figure}[tbh]
\includegraphics[width=0.45\textwidth]{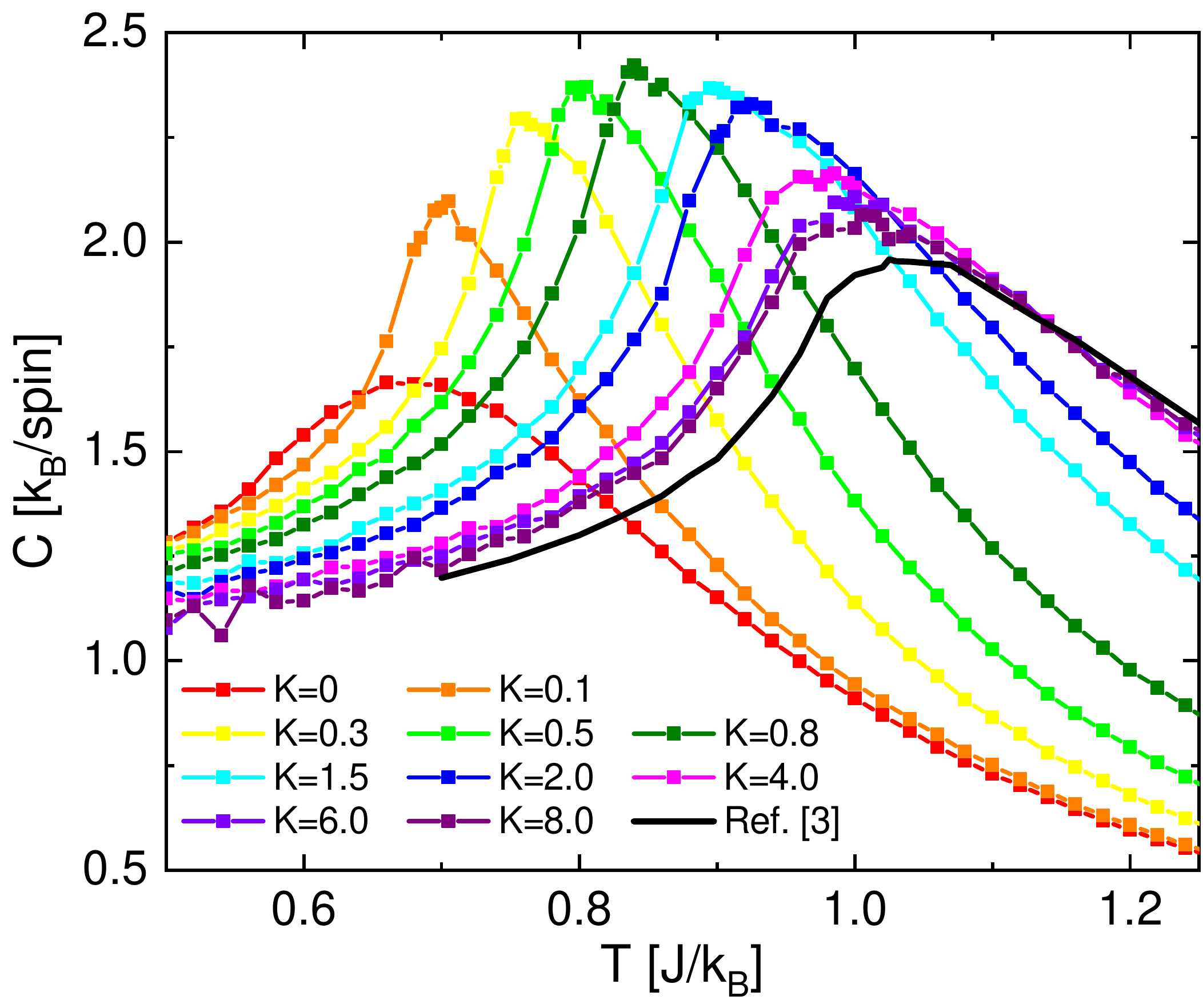}
\caption{specific heat (in units of $\kB$ per spin) as a function of temperature ($\kB T/J$) for different crystal-anisotropies $K$ (here in units of $J$). The black curve shows the result for the 2D $XY$ model published in Ref.\ \sbonlinecite{Gupta1992}, shifted by 1/2 because of the additional harmonic degree of freedom (compared to a true  $XY$ system). The spin waves contribute $\kB/2$ per degree of freedom to the specific heat.}
\label{fig:Kdep}
\end{figure}

When tuning the parameter $K$ from zero to large values, one can investigate the specific heat following different anisotropy strengths. For the isotropic case, $K = 0$, the specific heat features a hump, which is not associated with a phase transition. For a very large anisotropy $K \gg J$, the system's response is equivalent to that of a 2D $XY$ model, where the spin undergoes a BKT transition\cite{Gupta1992} at $\kB T/J = 1.04$. The specific heat curve reported by Gupta and co-workers in Ref.\ \sbonlinecite{Gupta1992} is shown in Fig.\ \ref{fig:Kdep} together with our simulation data.
Note that we have accounted for a constant shift of $\kB/2$ (per spin) between the \emph{true} $XY$ model and a very anisotropic Heisenberg model, as spin waves normal to the spin plane (always existing for Heisenberg models but absent in the $XY$ model) contribute a constant to the specific heat.

%

\section{Anisotropy of $\mathrm{Eu}^{2+}$ moments due to exchange interaction with $\mathrm{Fe}$ electrons}\label{app:crystal-anisotropy}
\begin{figure}[b!]
\includegraphics[width=0.48\textwidth]{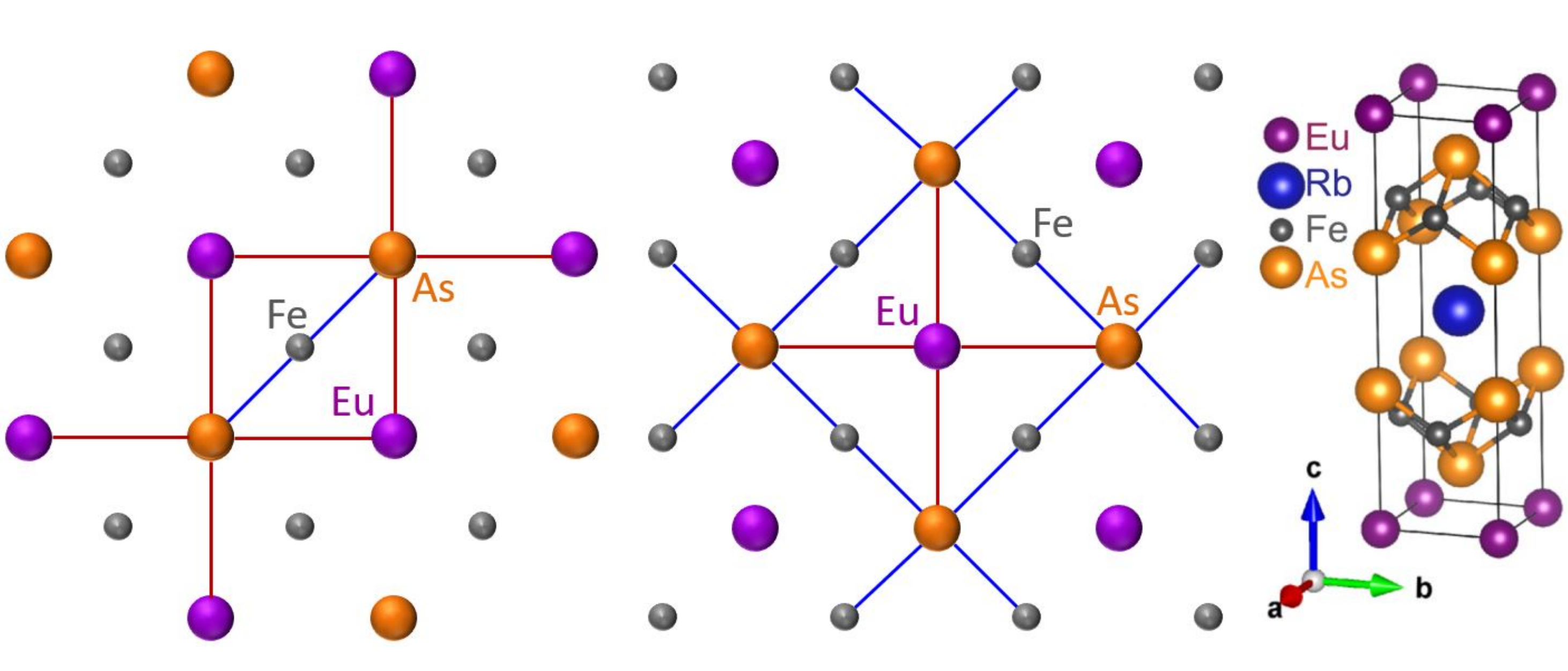}
\caption{Illustration of the superexchange interaction between $\mathrm{Eu}^{2+}$
spins and $\mathrm{Fe}$ $d$-electrons via $\mathrm{As}$ sites. Every
$\mathrm{Fe}$ $d$-electron interacts with 6 $\mathrm{Eu}^{2+}$ spins
per layer and every $\mathrm{Eu}^{2+}$ spin interacts with 12 $\mathrm{Fe}$
sites per layer.}
\label{fig:EuFeExchange} 
\end{figure}

The conventional mechanism of the crystalline anisotropy due to the
spin-orbital $\boldsymbol{L}\boldsymbol{S}$ interaction \cite{WhiteBook2007} is absent
for $\mathrm{Eu}^{2+}$ ions, because these ions do not have an orbital
moment $\boldsymbol{L}$ and their magnetic moment has purely spin
origin. We identify the interaction of the $\mathrm{Eu}^{2+}$ moments
with the $\mathrm{Fe}$ electrons as a possible source of crystalline
anisotropy as modeled in \eq \eqref{eq:anis2DHeisenberg}. To illustrate
this, we assume the exchange interaction between the $\mathrm{Eu}^{2+}$
spins $\vec{S}_{i}^{\mathrm{Eu}}$ and spins of electrons located
on $\mathrm{Fe}$ $d$-orbitals, $\vec{S}_{j}^{\mathrm{Fe}}$,
\begin{align}
\mathcal{H}_{\mathrm{Eu\text{-}Fe}}=-\sum_{\langle i,j\rangle}\tilde{J}_{ij}\vec{S}_{i}^{\mathrm{Eu}}\vec{S}_{j}^{\mathrm{Fe}}\label{eq:EuFeExchange}
\end{align}
with $\tilde{J}_{ij}$ being the Eu-Fe exchange constants. Assuming
that the mechanism behind this Eu-Fe interaction is superexchange
via $\mathrm{As}$ atoms, we observe that every $\mathrm{Fe}$ $d$-electron
interacts with 12 $\mathrm{Eu}^{2+}$ spins (6 per layer) while every
$\mathrm{Eu}^{2+}$ spin interacts with 24 $\mathrm{Fe}$ sites (12
per layer), see Fig. \ref{fig:EuFeExchange}. In the related compounds
without magnetic rare-earth layers, the response of the iron-arsenide
layers to an external field $\vec{H}$ is known to be anisotropic,
see Refs.\ \sbonlinecite{Wang2009} ($\mathrm{Ba}\mathrm{Fe}_{2}\mathrm{As}_{2}$)
and \sbonlinecite{MeierPhysRevB2016} ($\mathrm{Ca}\mathrm{K}\mathrm{Fe}_{4}\mathrm{As}_{4}$),
meaning that the local energy change of the iron subsystem caused
by the magnetic field $\vec{H}$ is $-[\chi_{x}(H_{x}^{2}+H_{y}^{2})+\chi_{z}H_{z}^{2}]/2$,
where $\chi_{x}$ and $\chi_{z}$ are the magnetic susceptibilities
per iron site. We should point out here that there are two possible contributions
to the anisotropy of the magnetic susceptibility. The magnetic moment
components of the iron $d$-electrons are related to their spin components
as $m_{\alpha}^{Fe}=g_{\alpha}\muB S_{\alpha}^{\mathrm{Fe}}$,
while the response of $S_{\alpha}^{\mathrm{Fe}}$ to the effective
magnetic field $h_{\alpha}=g_{\alpha}\muB H_{\alpha}$
is determined by spin susceptibility $S_{\alpha}^{\mathrm{Fe}}=\chi_{\alpha}^{\mathrm{spin}}h_{\alpha}$.
Both the $g-$factor and the spin susceptibility may be anisotropic due to
spin-orbit interactions and they both contribute to the anisotropy
of the magnetic susceptibility $\chi_{\alpha}=\left(g_{\alpha}\muB \right)^{2}\chi_{\alpha}^{\mathrm{spin}}$.
As follows from Eq.\ \eqref{eq:EuFeExchange}, the $\mathrm{Eu}^{2+}$
spins induce the effective field $\vec{h}_{\mathrm{Eu},j}\equiv\sum_{i}\tilde{J}_{ij}\vec{S}_{i}^{\mathrm{Eu}}$
on the iron subsystem yielding anisotropic nonlocal interaction of the form 
\begin{align}
\mathcal{H}_{\mathrm{Fe}} & =-\frac{1}{2}\sum_{\alpha,j}\chi_{\alpha}^{\mathrm{spin}}\Big(\sum_{i}\tilde{J}_{ij}S_{\alpha,i}^{\mathrm{Eu}}\Big) =-\frac{1}{2}\sum_{\alpha,j,i,k}\chi_{\alpha}^{\mathrm{spin}}\tilde{J}_{ij}\tilde{J}_{kj}S_{\alpha,i}^{\mathrm{Eu}}S_{\alpha,k}^{\mathrm{Eu}},
\end{align}
which extends over several neighboring $\mathrm{Eu}$ sites. We also
neglect here a possible nonlocality of iron-layer response. In the
regime when the correlation length of $\mathrm{Eu}^{2+}$ moments
exceeds the nonlocality range, we can approximate this interaction
by a local one. In this case an anisotropy in the $\mathrm{Eu}$ subsystem
is captured by the second term on the right-hand-side of \eq \eqref{eq:anis2DHeisenberg}
and we find an expression 
\begin{align}
K & =\frac{\tilde{J}_{\mathrm{eff}}^{2}}{4}\left(\chi_{x}^{\mathrm{spin}}-\chi_{z}^{\mathrm{spin}}\right)
=\frac{\tilde{J}_{\mathrm{eff}}^{2}}{4\muB ^{2}}\Big(\frac{\chi_{x}}{g_{x}^{2}}-\frac{\chi_{z}}{g_{z}^{2}}\Big)\label{eq:EffAnis}
\end{align}
 with the phenomenological constant with $\tilde{J}_{\mathrm{eff}}^{2}=\sum_{j,k}\tilde{J}_{ij}\tilde{J}_{kj}\left(S^{\mathrm{Eu}}\right)^{2}.$
If we denote the nearest-neighbor and next-neighbor exchange constants by $\tilde{J}_{1}$ and $\tilde{J}_{2}$ respectively then $\tilde{J}_{\mathrm{eff}}^{2}=32(\tilde{J}_{1}+2\tilde{J}_{2})^{2}(S^{\mathrm{Eu}})^{2}$.

It would be interesting to evaluate the Eu-Fe exchange constant $\tilde{J}_{\mathrm{eff}}$ from \eq \eqref{eq:EffAnis} and to compare it with the Eu-Eu exchange constant $J$. The reported susceptibilities for the iron moments in parent compounds, see Ref.\ \sbonlinecite{Wang2009} and \sbonlinecite{MeierPhysRevB2016} do not follow a Curie-Weiss law as expected for localized moments. Instead, for $\mathrm{Ba} \mathrm{Fe}_{2} \mathrm{As}_{2}$ \cite{Wang2009} the susceptibilities linearly increase with increasing temperature over a wide range ($150\text{-}400\mathrm{K}$). Furthermore the difference between the two susceptibilities (along and perpendicular to $c$) is almost temperature-independent and amounts approximately to $\chi_{x}-\chi_{z}\approx0.35\times10^{-3}\mathrm{emu/G\,mol}$. In the compound CaKFe$_{4}$As$_{4}$, which has the same structure as {\ooff}, the susceptibilities behave somewhat differently \cite{MeierPhysRevB2016}. In this case the susceptibilities linearly decrease with increasing temperature and their relative difference shrinks for higher temperature. Near the superconducting transition $\chi_{x}-\chi_{z} \approx 10^{-3}\mathrm{emu/G\,mol}$. Unfortunately, this information is not sufficient for an unambiguous evaluation of the Eu-Fe exchange constant from \eq \eqref{eq:EffAnis} because the $g$-factors of the $\mathrm{Fe}$ $d$-electrons remain unknown. If we make the simplest assumptions $g_{x}=g_{z}=2$ and $\tilde{J}_{2} \approx \tilde{J}_{1}/2$, we obtain $\tilde{J}_{1}\sim0.3J$, i.e. the Eu-Fe and Eu-Eu exchange strengths are comparable.

\vfill
\pagebreak
\twocolumngrid

\bibliographystyle{apsrev4-1-titles}
\bibliography{RbEuFe4As4}
\end{document}